\documentstyle[aaspp4,epsfig]{article}
 \def\simlt{\lower.5ex\hbox{$\; \buildrel < \over \sim \;$}}
  \def\simgt{\lower.5ex\hbox{$\; \buildrel > \over \sim \;$}}
\def\mag{\mbox{ mag}}
\def\kms{\mbox{ km s$^{-1}$}}
\def\mpc{\mbox{ Mpc}}
\def\kpc{\mbox{ kpc}}

\begin{document}
\title{A Fundamental Test of the Nature of Dark Matter}

\author{R. Benton Metcalf \& Joseph Silk\footnote{present address: Astrophysics, University of Oxford, Keble Rd. Oxford 0X1 3RH}}
\affil{\it Departments of Physics and Astronomy, and Center for Particle
Astrophysics \\ University of California, Berkeley, California 94720}

\begin{abstract}
Dark matter may consist of weakly interacting elementary particles or of macroscopic 
compact objects. 
We show that the statistics of the gravitational lensing of high redshift supernovae 
strongly discriminate between these two classes of dark matter candidates.  
We develop a method of calculating the magnification distribution of supernovae,  
which can be interpreted in terms of the properties of the lensing objects. 
With simulated data we show that $\simgt 50$ well measured type Ia supernovae 
( $\Delta m \sim 0.2\mag$ ) at redshifts $\sim 1$ can clearly distinguish macroscopic 
from microscopic dark matter if $\Omega_o \simgt 0.2$ and all dark matter is in one 
form or the other.
\end{abstract}


\section{Introduction}

The nature of dark matter (DM) poses one of the most outstanding problems in 
astrophysics. There are essentially two alternative hypotheses.  The dark 
matter may be microscopic, consisting  of weakly interacting particles (WIMPs) such as 
SUSY neutralinos or axions, or else be macroscopic, compact objects such as primordial 
black holes (PBHs), brown dwarfs or old white dwarfs (MACHOs).  Big Bang
nucleosynthesis (BBN) puts a bound on the density in baryonic matter of $\Omega_b h^2 
\simlt 0.02$ (or $\simlt 0.03$ if one allows for inhomogeneous BBN), 
but the density of PBHs is not well constrained.  It is possible that some hitherto 
unknown mechanism allows for DM that is dominated by macroscopic objects.  For these 
reasons direct observational constraints on macroscopic DM of any density are very 
important.

We propose a simple test for distinguishing macroscopic from microscopic dark 
matter. In this letter we consider only the opposing hypotheses that one or the other 
dominates.  If the DM is microscopic, the clustered component, in halos, lenses 
high redshift supernovae (SNe).  If the DM is macroscopic, most light beams do not 
intersect any matter - no Ricci focusing - and the SN brightness distribution is 
skewed to an extent that can be quantitatively distinguished from halo lensing.  

\section{Properties of the Magnification Probability Distribution Function}
\label{pdfsection}

In this paper we  consider the lensing of distant supernovae by discrete ``lenses''.  
A lens is the smallest unit of mass that acts coherently for the purpose of lensing.  
This could be a galaxy halo or it could be a high mass dark matter candidate such as a PBH. 

We make the distinction between macroscopic and microscopic DM more quantitative 
by considering two mass scales.  The first is defined by the requirement that the projected 
density be smooth on the scale of the angular size of the source.  This gives a maximum mass of
\begin{eqnarray}
m_s \sim 6\times 10^{10}\mbox{g}~ \left( \frac{\lambda_{s}}{\mbox{AU}} \right)^3 \Omega_o h^2 f
\end{eqnarray}
where $\lambda_s$ is the physical size of the source and $f$ is a geometric factor of order 
unity.  If the unit of DM is smaller then this it is microscopic DM.  Another, larger mass 
scale is defined by the requirement that the angular size of the source be small compared 
to the Einstein ring radius so that it can be considered a true point source if
\begin{eqnarray}
m \simgt 5\times10^{-7}\mbox{M}_\odot~ \left( \frac{10^3 \mbox{Mpc}}{D_s} \right) \left( \frac{\lambda_s}{\mbox{AU}} \right)^2 f
\end{eqnarray}
where $D_s$ is the angular size distance to the source.  If a lens is much below this mass 
the high magnification tail of the distribution function will be suppressed and the rare high magnification events will become time-dependent(\cite{SW87}).
The measured velocity of the expanding photosphere of a type Ia SN is around 
$1.0-1.4 \times 10^4 \kms$ (\cite{Patat96})which means $\lambda_s \sim 40-57 \, \Delta t \mbox{ AU}/\mbox{week}$.  The SN reaches maximum light in approximately one week.

The background cosmology will be taken to be the standard 
Friedman-Lema\^{\i}tre-Robertson-Walker (FLRW) with the metric 
$ds^2 = dt^2 +a(t)^2\left(d\chi^2 + D(\chi)^2 d \Omega\right)$ where the comoving angular 
size distance is $D(\chi) = \{ R\sinh(\chi/R),\chi,R\sin(\chi/R) \}$ ($R=|H_o 
\sqrt{1-\Omega_o-\Omega_{\Lambda}}|^{-1}$) for the open, flat and closed global geometries respectively.  Another relevance angular size distance is the Dyer-Roeder or empty-beam 
distance, $\tilde{D}(\chi)$ (\cite{DR74,Kant98}, note difference in notation) which is angular size distance for a beam that passes through empty space and experiences no shear.

\subsection{Magnification by a single lens}
\label{singlepdf}

Consider a single lens at a fixed coordinate distance from Earth.  The path of the light is 
described by either of two lensing equations:
\begin{eqnarray}
{\bf r}_\perp = {\bf y} - {\bf \alpha}({\bf y},\tilde{D}_l,\tilde{D}_s) \label{lenseq1}
\\
{\bf r}_\perp = [1-\kappa_b(\chi_s)]{\bf y} - {\bf \alpha}({\bf y},D_l,D_s) \label{lenseq2}
\end{eqnarray}
where ${\bf r}_\perp$ is the positions of the lens relative to the undeflected line of 
sight to the source, ${\bf y}$ is the position of its image in the same plane and ${\bf \alpha}$ is the deflection angle times the angular size distance.  
In equation~(\ref{lenseq1}) a negative background convergence, $\kappa_b$, is 
included to account for the lack of background mass density that is assumed when $D$ 
is used instead of $\tilde{D}$.  
Two magnifications, $\tilde{\mu}$ and $\mu$, can be defined using equations~(\ref{lenseq1}) 
and (\ref{lenseq2}) respectively.  
The requirement that the two lensing equations agree on the true size of an object results in the relation $\tilde{D}(\chi)=[1-\kappa_b(\chi) ] D(\chi)$.  The explicit form of $\kappa_b(z)$ can be found by comparing the standard FLRW expression for $D(\chi)$ with the solutions for $\tilde{D}(\chi)$ found in \cite{Kant98}.  

The probability that the lens is located between $r_\perp$ and $r_\perp + dr_\perp $ is 
$p(r_\perp)dr_\perp \propto r_\perp dr_\perp$.  
If the lens is spherically symmetric and the magnification is a monotonic function of 
$r_\perp$ the expression for the magnification can be inverted (at least numerically) to 
get $r_\perp(\mu,D,D_s)$.  Then probability of a lens causing the magnification $1+\delta\mu$ can be found by changing variables.  Lenses might also have properties such as mass, scale length, etc. which need to be averaged.

For the case of a point mass lens the total magnification of both images is given by $\tilde{\mu}=\hat{r}^2 +2/\hat{r}\sqrt{\hat{r}^2 +4}$; $\hat{r} \equiv r_\perp/R_e(m,D,D_s)$.
The Einstein radius of the lens is given by $R_e^2= 4Gm \tilde{D}_l \tilde{D}_{ls}/\tilde{D}_s$.
The single lens distribution function is then
\begin{equation}
p(\delta\tilde{\mu}) d\delta\tilde{\mu} \propto \left[ (1+\delta\tilde{\mu})^2 -1\right]^{-3/2} d\delta\tilde{\mu}.
\label{pointprob}
\end{equation}
The probability in (\ref{pointprob}) is not normalizable; 
it diverges at small $\delta\tilde{\mu}$.  
This can be handled by introducing a cutoff in either $\delta\tilde{\mu}$ space or in 
$r_\perp$.  The nature of this cutoff is not important as long as it is at sufficiently 
small $\delta\tilde{\mu}$ or large $r_\perp$.  This will be clear when the total 
magnification distribution due to multiple lenses is considered.

If the dark matter consists of microscopic particles clumped into halos, the entire halo 
will act as a single lens.  In this case the Ricci focusing contribution to the 
magnification strongly dominates over shear distortions produced by mass outside of the beam (\cite{HW98,PMM98}) and is then a function of only the local dimensionless surface 
density, $\kappa({\bf y})$.  Furthermore the lensing of the great majority of SNe will be 
quite weak which allows us to confidently make the linear approximation: $\delta\mu = 
2[\kappa({\bf y},D_l,D_s)+\kappa_b]$.  This assumption has been well justified by 
many authors and will be confirmed by results in \S\ref{total}.

For the purposes of this paper it will suffice to use a simple model for the surface 
density of halos.  We use models with surface densities given by
\begin{equation}
\Sigma(y_\perp) = \frac{V_c^2}{2Gy_\perp} \left[ \left(\frac{y_\perp}{r_s}\right)^2 +1 \right]^{-1}
\label{massdens}
\end{equation}
This model behaves like a singular isothermal sphere out to $y_\perp \simeq r_s$ where it is 
smoothly cut off.

In the following calculations, each halo is assumed to harbor a galaxy.  At all redshifts 
a Schechter luminosity function fit to local galaxies is assumed with $\alpha=-1.07$ and $\phi^*=0.01(1+z)^3h^3\mpc^{-3}$.  The luminosities 
are then related to the circular velocity, $V_c$, by the local Tully-Fisher relation, 
$V_c=V_*(L/L_*)^{0.22}$ where $V_*=200\kms$.  The scale lengths are related to 
the luminosity through $r_c=r_*(L/L_*)^{1/2}$ with $r_*=220\kpc$.  The precise values used for these 
parameters do not have a significant effect on the results of this paper.

\subsection{Total Magnification}
\label{total}

The total magnification of a source includes contributions from all the lenses surrounding 
the light path.  To find the true path connecting a source to us, the lensing equation must be 
solved with multiple deflections (see \cite{sch92}).  The magnifications due to different lens 
planes are in general nonlinearly coupled.  However, if the deflections due to no 
more than one of the lenses is very weak the coupling between lenses can be ignored and their 
magnifications, $\delta\mu$ or $\delta\tilde{\mu}$, will add linearly.  This 
is a good approximation for the vast majority of light paths in realistic models.  
The validity of this assumption will be justified by the results and is further investigated 
in \cite{Met99}.  
Furthermore numerical simulations and analytic arguments show that for both kinds of DM it is 
a good approximation to take the lenses to be uncorrelated in space (see \cite{HW98,Met98b}). 
If in addition we take the lenses internal properties to be uncorrelated, the probability 
that the total magnification, $\delta\tilde{\mu}_s$, 
of a point source is between $\delta\tilde{\mu}_s$ and $\delta\tilde{\mu}_s+d\delta\tilde{\mu}_s$ is
\begin{eqnarray}
P(\delta\tilde{\mu}_s)d\delta\tilde{\mu}_s =d\delta\tilde{\mu}_s~ \int \prod_{i=1}^N\left[d\delta\tilde{\mu}_i \; p(\delta\tilde{\mu}_i)\right] \nonumber \\ 
\delta(\delta\tilde{\mu}_s-\sum_{i=1}^N \delta\tilde{\mu}_i)
\label{Prob}
\end{eqnarray}
where the $\delta\tilde{\mu}_i$ is the contribution of the $i$th lens. 

The magnification $\delta\mu_s$ is defined as the deviation of the luminosity 
from its mean value.  As a result, the mean of the distribution $P(\delta\mu_s)$ must 
vanish.\footnote{The actual mean angular size distance should be slightly larger than the 
FLRW value because galaxies obscure some sources.  Galaxies are presumably correlated with 
high density regions through which the magnification would be above average were they 
transparent.}
This combined with the requirement that both magnifications agree on the true size of a source 
results in the expression $1-\kappa_b(\chi)=\langle{\tilde\mu}(\chi)\rangle^{1/2}$.  In this 
way the value of $\kappa_b(\chi)$ can be found by calculating the mean of (\ref{Prob}) 
numerically and a consistency check of the calculations can be made by comparing the results 
with the explicit values for $D(\chi)$ and 
$\tilde{D}(\chi)$.  These values agree to a few percent which is consistent with the 
uncertainty introduced by the discrete nature of the numerical calculation in the power law 
tail of the distribution.
The minimum magnification, $\delta\mu_{min}$, in the single lens distribution is set low 
enough that the resulting total distribution is independent of the cutoff.

\begin{figure}[t]
\centerline{\epsfysize=7 cm{\epsfbox{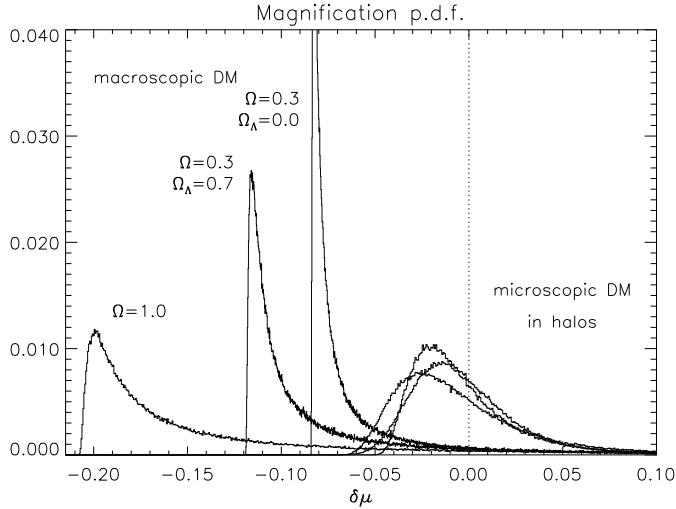}}}
\caption[stuff]{Histograms representing the total magnification probability distribution for macroscopic DM and microscopic DM clumped into halos.  The means of all the distributions are zero.  For the macroscopic DM case all the matter in the universe is in the lenses.  The shape of the distribution for DM halos is dependent on both the cosmology and the specific halo model assumed.  This is a representative sample.
\label{pdf}}
\end{figure}

Figure~\ref{pdf} shows some examples of histograms made by producing random values 
$\delta\tilde{\mu}_i$ drawn from the single lens distributions and then adding them to get 
the total magnification.  The macroscopic DM distributions shown in 
figure~\ref{pdf} are independent of the lens mass and peak well below their mean and near 
the empty-beam solutions (corresponding to $\delta\mu=-0.21$, $-0.12$ and $-0.084$) because 
in these cases most lines of sight do not come very close to any lens.  
The probability that there are two lenses which individually give magnifications greater than 
$\delta\mu$ becomes appreciable only below the peak.  This supports our approximation that 
whenever the lensing is strong it is dominated by one lens and the coupling between lenses is 
small at this redshift.  In addition we have compared our results with the numerical 
simulations of \cite{HW98} and found excellent agreement.

\section{Distinguishing Dark Matter Candidates}
\label{statsec}

The apparent luminosity of a SN, $l_{ob}$, after lensing can be expressed in terms of either 
of the two magnifications, $l_{ob} = \mu \, l =\tilde{\mu} \, l/\langle\tilde{\mu}\rangle$.
We wish to infer via the measured luminosities of a set of SNe, each located at a different 
redshift, from which distribution the magnifications were drawn and in this way surmise 
which DM candidate is most likely.  To establish some insight into the magnitude of this 
effect, the differences in magnitudes between the average and the empty-beam solutions at 
$z=1$ are: $-0.25\mag$ for $\Omega_o=1$, $-0.14\mag$ for flat $\Omega_o=0.3$ and $-0.10\mag$ 
for open $\Omega_o=0.3$.

Let us denote the probability of getting a data set, $\{\delta\mu\}$, given a model 
- either microscopic or macroscopic DM - as $P(\{\delta\mu\} | model)=\prod_i P(\delta\mu_i | model)d\delta\mu_i$ where the product is over the observed SNe.  The model here includes sources of noise.  
This probability can be calculated numerically from the probability distributions discussed in \S\ref{total}.  Because of Bayes' theorem we 
know that the ratio of these two probabilities is equal to the relative likelihood of the 
models being correct, the odds, given a data set.  It is convenient to modify the odds 
into the statistic
\begin{eqnarray}
{\cal M}_p\equiv \frac{1}{N_{SN}} \ln\left[\frac{ \int d\Omega_o d\Omega_\Lambda p(\Omega_o,\Omega_\Lambda) P\left(\{\delta\mu\} | \mbox{ macroDM,noise}\right)}{\int d\Omega_o d\Omega_\Lambda p(\Omega_o,\Omega_\Lambda) P\left(\{\delta\mu\} | \mbox{ halos,noise}\right)}\right].
\label{Mp}
\end{eqnarray}
where $p(\Omega_o,\Omega_\Lambda)$ is the prior distribution for the cosmological model 
based on independent information or prejudice.  The measured ${\cal M}_p$ is expected to 
be large if DM is macroscopic and smaller if DM is microscopic or nonexistent. 

\begin{figure}[t]
\vspace{-3cm}
\centerline{\epsfysize=12 cm{\epsfbox{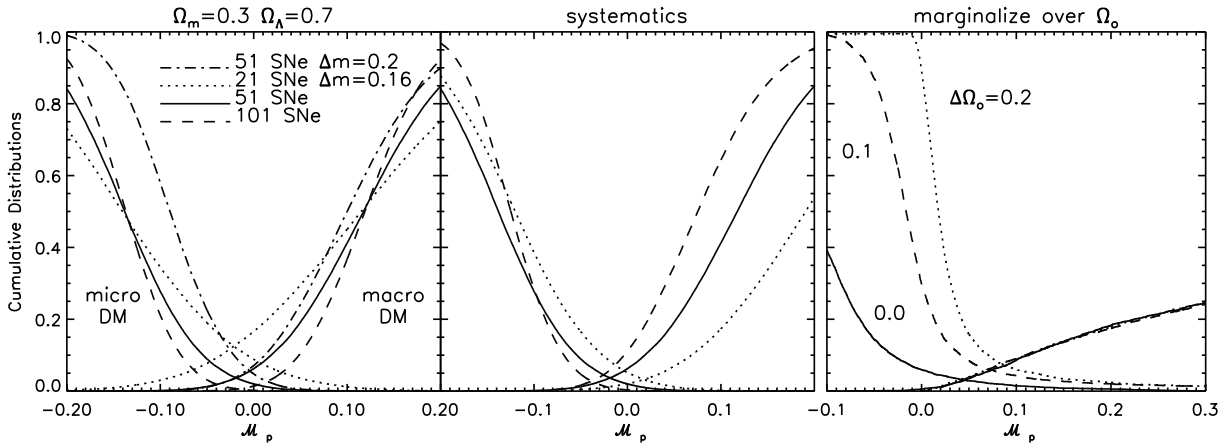}}}
\vspace{-3cm}
\caption[stuff]{Differentiating dark matter candidates: The cumulative distributions of the statistic ${\cal M}_p$.  The cases where the true DM is macroscopic rise toward the right and microscopic DM cases rise toward the left.  All the observed SNe are taken to be at $z=1$ and in all cases $\Omega_o+\Omega_\Lambda=1$.  On the left the different curves are for different numbers of observed SNe as marked with $\Delta m=0.16\mag$ in all cases except the dot-dashed curves.  The halos have total density $\Omega_h=0.27$.  The sold curves in the center panel are the same as in the left hand panel and the other curves are described in the text.  In the right hand panel there are 75 SNe observed.  The widths of the prior distributions are marked.%$\Delta\Omega_o=0$ (solid), $0.1$ (dashed) and $0.2$ (dotted).
\label{bayes}}
\end{figure}

For the left hand plot in figure~\ref{bayes} five thousand simulated data sets were created, 
${\cal M}_p$ is calculated for each of them and their cumulative distributions plotted.  
The noise included in the simulation originates from both the intrinsic 
distribution of SN luminosities, presently corrected to $\sim 0.12\mag$, and the 
observational noise, presently an additional $\sim 0.08\mag$.  For the left hand plot the 
noise is taken to be Gaussian-distributed in magnitudes with a standard deviation of 
$0.16 \mag$ except for the dot-dashed curves which have $\Delta m =0.2\mag$.  The cosmology 
is fixed in this plot, ie $p(\Omega_o,\Omega_\Lambda)$ is a $\delta$-function.  
${\cal M}_p$ can be calculated for a given data set and compared to this plot to determine 
its significance.  It can be seen here that for 51 SNe (solid curve) 
at $z=1$ the two distributions overlap at the $4\%$ level, ie $96\%$ of the time one of 
the DM candidates can be ruled out at better than the $96\%$ confidence level.
One of the advantages of ${\cal M}_p$ is that it is close to Gaussian distributed with 
a mean that is independent of the number of SNe observed.  In this way, once the cosmology 
and noise model is fixed, the value of ${\cal M}_p$ is a direct prediction of the kind of DM.

The middle plot in figure~\ref{bayes} illustrates the importance of some possible systematic 
uncertainties that arise from not knowing precisely the distribution of the noise.  The 
solid curves are the same as in the left hand plot.
The dotted curve is the extreme case where the noise is actually Gaussian distributed in 
magnification (there is a low magnitude tail), but ${\cal M}_p$ is calculated under the same 
assumptions as in the left hand plot.  The dashed line in this plot is the case where 
the standard deviation is overestimated to be $\Delta m =0.2\mag$ but is really $\Delta m 
= 0.16\mag$.  These errors in the noise model do not destroy the efficacy of the test, but 
they could be important if a long tail exists in the intrinsic distribution of luminosities 
and they become more important for smaller $\Omega_o$ and $\Omega_\Lambda$.

The right hand plot in figure~\ref{bayes} addresses the question of differentiating between 
DM candidates without assuming specific values for the cosmological parameters, thereby making 
the conclusion cosmology independent.  Here the prior is taken to be $p(\Omega_o,\Omega_\Lambda)=\delta(1-\Omega_o-\Omega_\Lambda)$ 
within a range in $\Omega_o$ ($\Delta\Omega_o=0$, $0.1$ and $0.2$) centered on $0.3$ and 
zero otherwise.  The simulated data is the same here as for the solid curves in the two 
left hand plots.  However, the integrations in (\ref{Mp}) would be prohibitively time 
consuming if the entire magnification distribution function were calculated for each trial 
cosmology.  To simplify the calculation without loosing much of the test's effectiveness 
we use approximate, analytic test distribution functions.  For the macroscopic DM case we 
use (\ref{pointprob}) with the low magnification cutoff which insures that it gives the 
correct mean.  Comparison of this approximation with the full multi-lens distribution shows 
that it is a good approximation especially for low $\Omega_o$.  For the microscopic DM/halo 
case we approximate the distribution as a Gaussian with an appropriate width 
(see \cite{Met98a}).  This plot shows that not only is this simplified calculational 
technique adequate, but that one does not need to assume a precise cosmological model to 
differentiate between DM candidates.  Increasing the width of the prior beyond 
$\Delta\Omega_o=0.2$ does not make much difference.  The reason for this is that if the 
assumed cosmological parameters are significantly different than the true ones the 
distribution will be shifted to an extent that it is no longer consistent with the data.  
This shift would be confused for a lensing effect if the two kinds of distributions, 
illustrated in figure~\ref{pdf}, were translations of each other, but they are not, even 
after noise is added.  For the two DM cases, the modes of the magnification distributions 
follow different $m$--$z$ relations, but their means are the same.  For a fixed redshift, it is 
the distribution of luminosities about the mean that distinguishes the two cases.

For open models ($\Omega_\Lambda =0$) it is more difficult to differentiate the DM candidates, 
but even in this case with 51 SNe at $z=1$ and $\Omega_o=0.3$ we expect to get better 
than $90\%$ confidence at least $90\%$ of the time.  If $\Omega_o=0.1$ BBN constraints just 
barely allow for all DM being made of baryonic objects.  In this case similar bounds to those 
shown in figure~\ref{bayes} for 51 SNe can be achieved with 200 SNe.  However the means of the 
${\cal M}_p$ distributions are closer together in this case, making the test more susceptible 
to systematic errors in the assumed noise model.

The power of lensing to differentiate DM candidates comes mostly from its ability to 
identify macroscopic DM.  A positive detection of the lensing by microscopic DM halos will 
take more SNe, as will constraining the precise fraction DM in macroscopic form, unless 
correlations between SN luminosities and foreground galaxies are utilized 
(\cite{Met98b,Met99}).

\section{Discussion}

One concern in implementing the test described here is the possibility that type Ia SNe 
and/or their galactic environments evolve with redshift.  This is also a major concern in 
cosmological parameter estimation from SNe.  So far there is no indication that the colors
 or spectra systematically change with redshift (\cite{Perl97}, \cite{Riess98}).  
Since the evolution of the magnification distribution is determined by cosmology it is in 
principle possible to make an independent test for systematic evolution in the distribution of SN luminosities.  

Microscopic DM does not need to be clustered for this test to 
work.  The clustering is added to make the calculations realistic.  Clustering the 
microscopic DM to a greater or lesser extent would affect our results quantitatively, but the 
test would still be viable in more extreme cases.
We conclude that if the assumptions we have made about the noise levels in future SN 
observations remain reasonable,  on the order of $50 - 100$ SNe 
at $z\sim 1$ should suffice  to determine a fundamental question:
whether the major constituent of extragalactic DM 
is microscopic particles or macroscopic objects.

\acknowledgments
We would like to thank D. Holz for providing the results of his simulations for 
the purposes of comparison.

\end{document}